\documentstyle[preprint,aps,prl]{revtex}

\begin{document}

\draft

\title{ Microscopic calculation of the phonon-roton branch in
superfluid $^{\bf 4}$He}

\author{ J. Boronat and J. Casulleras }

\address{Departament de F\'{\i}sica i Enginyeria Nuclear,
Campus Nord B4-B5, \protect\\ Universitat Polit\`ecnica de Catalunya,
E-08028 Barcelona, Spain}

\date{July 29, 1996}

\maketitle

\begin{abstract}
Diffusion Monte Carlo results for the phonon-roton excitation branch in
bulk liquid $^4$He at zero temperature are presented. The sign problem
associated to the excited wave function has been dealt both with the
fixed-node approximation  and the released-node
technique. The upper bounds provided by the fixed-node
approximation are shown to become exact when using the released-node
method. An excellent
agreement with experimental data is achieved both at the equilibrium
and near the freezing densities.
\end{abstract}


\narrowtext

%\twocolumn

The physical nature of the excitations in superfluid $^4$He at low
momenta is still nowadays not completely well understood
\cite{gly1}, in contrast
with the vast knowledge of its static properties. In a pioneering work,
Landau \cite{landau} proposed a model for the elementary excitations to
explain
the superfluidity of liquid $^4$He. From a different point of view,
Bogoliubov \cite{bogo} calculated the excitation spectrum of a weakly
interacting
Bose gas where the condensate fraction, i.e., the fraction of particles
in the zero momentum state, plays an explicit role. Both theories
predict
a continuous dispersion curve which starts with a phonon excitation,
reaches a first maximum (maxon), lowers to a local minimum (roton), and
then grows up as the energy of a free particle. From general ideas on
the nature of the excitations in an interacting Bose fluid, Feynman
\cite{fey1}
proposed the first microscopic approach to the problem. The Feynman
trial wave function provides a qualitative description of the excitation
spectrum but fails in reproducing the roton energy by a factor two.
Later on, Feynman and Cohen \cite{feyco} included backflow
correlations in the trial wave function,
reducing in one half the differences
between the experimental data and the original Feynman's prediction.
Following  Feynman's language, the phonon-roton branch corresponds to
collective density-like excitations where the condensate fraction does
not enter in an explicit way. Recently, it has been argued by Glyde and
Griffin \cite{gly2} that the continuous spectrum results from the
superposition
of density excitations, dominating in the phonon region, and
single-particle excitations, important in the roton minimum. This theory
has emerged after the experimental determination of the temperature
dependence of the excitation spectrum which shows
the phonon peak in
the dynamic structure function $S(q,\omega)$  at both sides
of the $\lambda$-transition whereas the roton practically disappears in
the normal phase.

In the last years a considerable effort has been made to improve
quantitatively the microscopic predictions for the phonon-roton
excitation spectrum $\varepsilon(q)$. Manousakis and Pandharipande
\cite{pandha}
calculated $\varepsilon(q)$ by means of the correlated basis function
(CBF) method using a basis of Feynman-Cohen states.
The variational Monte
Carlo (VMC) method using shadow wave functions has also proved to be
quantitatively quite efficient in the calculation of $\varepsilon(q)$
\cite{shadow} in spite of its approximate
description of the ground state. The application of {\it ab initio}
Monte Carlo methods to this problem has been, however, severely
hindered by the sign problem associated to the excited wave function.
Only recently, Boninsegni and Ceperley \cite{boni} have calculated
$\varepsilon(q)$
by means of a path integral Monte Carlo (PIMC) calculation of
$S(q,\omega)$ from a Laplace inversion of the
imaginary-time correlation factor $S(q,t)$. However, for noisy
data this inversion is an ill-posed problem that prevents a
model-independent determination.

In the present work, we present a zero temperature calculation of the
phonon-roton spectrum using the diffusion Monte Carlo (DMC) method. In
this method, the imaginary-time Schr\"odinger equation for the function
$f({\bf R},t) = \psi_T({\bf R})\Phi({\bf R},t)$,
\begin{equation}
-\frac{\partial f({\bf R},t)}{\partial t} = -
D\nabla_{\bf R}^2 f({\bf R},t) + D\nabla_{\bf R}({\bf F}({\bf R})f({\bf
R},t)) + (E_L({\bf R}) - E) f({\bf R},t) \ ,
\label{dmc}
\end{equation}
is solved stochastically, $\Phi({\bf R},t)$ and $\psi_T({\bf R})$
being the wave function of the
system and  a trial wave function used for importance
sampling, respectively.  In the above equation, $E_L({\bf
R})=\psi_T^{-1}({\bf
R})H\psi_T({\bf R})$ is the local energy and ${\bf F}({\bf
R})=2\psi_T^{-1}({\bf R})\nabla_{\bf R} \psi_T({\bf R})$ acts as a
quantum drift force; $D=\hbar^2/2m$, with $m$ the mass of the
particles, and ${\bf R}$ stands for
the $3N$-coordinate vector of the $N$ particles. In the asymptotic
regime, $f({\bf R},t \rightarrow \infty) \longrightarrow
\psi_T({\bf R}) \, \Phi({\bf R})$ where $\Phi({\bf R})$ corresponds to
the lowest energy eigenstate of the system not orthogonal to the trial
wave function $\psi_T({\bf R})$. For a thorough  description of the DMC
method see, for instance, Refs. \cite{rey,boro}.

The wave function $\psi({\bf R})$ corresponding to a phonon-roton
excitation is an eigenstate of the momentum operator. As pointed
out by Feynman \cite{fey1}, this requirement is achieved with the simple
model wave function
\begin{equation}
\psi^F ({\bf R}) = \sum_{i=1}^{N} e^{i {\bf q} \cdot {\bf r}_i} \:
\psi^0({\bf R})  \ ,
\label{fey1}
\end{equation}
$\psi^0({\bf R})$ being the ground-state wave function. The first
correction to $\psi^F ({\bf R})$, originally proposed by Feynman and
Cohen \cite{feyco}, includes backflow correlations.
In this case, the
excited wave function is given by
\begin{equation}
\psi^{BF} ({\bf R}) = \sum_{i=1}^{N} e^{i {\bf q} \cdot \tilde{ {\bf
r}}_i} \: \psi^0({\bf R})  \ ,
\label{back1}
\end{equation}
where
\begin{equation}
 \tilde{ {\bf r}}_i = {\bf r}_i + \sum_{j \neq i}  \eta(r_{ij}) \: {\bf
r}_{ij}   \ .
\label{backr}
\end{equation}
The inclusion of backflow correlations improves appreciably the
variational results of $\varepsilon(q)$ with respect to the Feynman's
choice
but the quantitative agreement with experiment is still poor, specially
near the roton minimum.

In the DMC implementation a real probability distribution
function $f({\bf R},t)$ is suitable. Therefore, we choose as importance
sampling
wave function $\psi_T({\bf R})$ the superposition of two excitations of
momenta ${\bf q}$ and $-{\bf q}$ which are degenerate in energy:
\begin{equation}
\psi^F_T ({\bf R}) =  \sum_{i=1}^{N} \cos ( {\bf q} \cdot {\bf
r}_i)  \: \psi^0_T({\bf R})
\label{fey2}
\end{equation}
at Feynman's level, and
\begin{equation}
\psi^{BF}_T ({\bf R}) =  \sum_{i=1}^{N} \cos ( {\bf q} \cdot
\tilde{ {\bf r}}_i)  \: \psi^0_T({\bf R})
\label{back2}
\end{equation}
when backflow correlations are included.

In a first step, the sign problem associated to the excited wave
function has been dealt within the framework of the fixed-node (FN)
approximation \cite{rey}. The FN approximation provides an upper bound
to the exact
value and has been extensively used in the study of fermionic systems.
In what concerns the calculation of excited states, we have recently
used the FN-DMC method to study a vortex excitation in two-dimensional
superfluid $^4$He \cite{giorgi}. Within FN, the sign problem is avoided
by imposing
the nodal surface of the trial wave function to the excited Monte
Carlo wave function.
The problem is hence mapped onto a bosonic calculation
with a well defined density probability function. Beyond this
approximation, one can remove the
nodal constraint of the FN approach using a released-node technique (RN)
\cite{ceper}. This method makes use of an auxiliary guiding
function $\psi_g({\bf R})$, positively defined everywhere. Random
walkers are then allowed to cross the nodal surface (becoming
negative) and to survive for a finite lifetime $t_r$. The RN approach
provides the exact eigenvalue in the limit of large surviving times, but
presents the disadvantage of becoming
numerically unstable in the limit $t_r \rightarrow \infty$ where the
number of positive and negative walkers become similar.

The success of the method,
i.e., the achievement of an asymptotic regime before the growth of the
statistical errors, is closely related to the quality of $\psi_T({\bf
R})$ and $\psi_g({\bf R})$.
The guiding function $\psi_g({\bf R})$
has to approach $|\psi_T({\bf R})|$  away from the
nodal surface and must be non-zero in the nodes to make possible the
flux of walkers through it. We have taken the simple  model
\begin{equation}
\psi_g({\bf R})= \left( \psi_T({\bf R})^2 + a^2 \right)^{1/2} \ ,
\label{guia}
\end{equation}
which satisfies both requirements for a proper choice of the value of
the parameter $a$. In fact, the value given to the parameter $a$, which
has to be of the same order of magnitude than the mean value of
$|\psi_T({\bf
R})|$, governs the flux of walkers through the nodes. Therefore, the
relaxation time of the release process is a function of $a$: when the
value
of $a$ is increased the relaxation time is  reduced and vice versa.

The released-node energy is obtained projecting out the excited state
modeled by $\psi_T({\bf R})$. This projection is carried out assigning
to each walker a weight $W({\bf R})$ given by
\begin{equation}
  W({\bf R})= \sigma({\bf R})  \: \frac{|\psi_T({\bf R})|}{\psi_g({\bf
R})}      \ ,
\label{pes}
\end{equation}
$\sigma({\bf R})$ being $+1$ ($-1$) for an even (odd) number of
crossings. The released-node energy is thus determined through
\begin{equation}
 E_{RN}(t_r) = \frac{\sum_{t \leq t_r} W({\bf R}) \: E_L^T({\bf R})}
               {\sum_{t \leq t_r} W({\bf R})}    \ ,
\label{rnener}
\end{equation}
where the sums are extended to all the surviving walkers with a lifetime
less than $t_r$, and $E_L^T({\bf R})=\psi_T({\bf R})^{-1} H \psi_T({\bf
R})$.

In our computation we have considered a simulation box with 108 $^4$He
atoms interacting through the HFD-B(HE) Aziz potential \cite{aziz}, and
at two
densities, $\rho_0=0.365\ \sigma^{-3}$ and $\rho_P=0.438 \ \sigma^{-3}$
($\sigma=2.556$ \AA). The density $\rho_0$ corresponds to the
equilibrium density, and $\rho_P$ is close to the freezing density. The
ground-state correlations have been modeled by a two-body wave function
originally proposed by Reatto \cite{reat},
\begin{equation}
\psi_T^0({\bf R}) = \prod_{i<j} \exp \left\{ -\frac{1}{2} \left(
\frac{b}{r_{ij}} \right)^5 - \frac{L}{2} \exp \left[ - \left(
\frac{r_{ij}-\lambda}{\Lambda} \right)^2 \right] \right\}  \ ,
\label{reat}
\end{equation}
which we have previously employed in ground-state calculations
\cite{boro}. The
function $\eta(r)$, entering in the backflow wave function
(\ref{backr}), has been chosen to be a gaussian
\begin{equation}
  \eta(r)= A \: \exp \left[ - \left( \frac{r-r_b}{\omega_b} \right)^2
\right]
\label{gauss}
\end{equation}
as in the variational
calculation of Ref. \cite{pandha}. The values of the parameters in Eqs.
(\ref{reat},\ref{gauss})
are $b=1.20\ \sigma$, $L=0.2$, $\Lambda=0.6\ \sigma$,
$\lambda=2.0 \ \sigma$, $A=0.15$, $r_b=0.8\ \sigma$, and $\omega_b=0.44
\ \sigma$, which are roughly the optimal ones at the equilibrium density
$\rho_0$. The parameter $a$ appearing in the guiding wave function
$\psi_g({\bf R})$ (\ref{guia}) has been taken as $a=3.0$ for all the $q$
values, and the largest lifetime $t_r$ used corresponds to
 190 DMC sweeps. The same set of parameters have been used at the
highest density $\rho_P$.

As mentioned before, the released-node energy estimation would be exact
in the
limit of large lifetimes. However, the computational
effort to
simultaneously enlarge $t_r$  and maintain the statistical
fluctuations into an acceptable level grows with $t_r$.
Having estimated a reasonable  upper limit of $t_r$, given the
present computational resources,
we can study the influence of the excited trial wave
function in the RN energy. As a general trend, if the RN energy does not
reach a constant regime within $t_r$, an improved model for
the excited
trial wave function should be used. The empirical way in which we
have studied the asymptotic regime of the RN energy is by fitting the
function
\begin{equation}
  E (t_r) = E_{\infty} + C \: e^{-t_r / \tau}
\label{fit}
\end{equation}
to the largest $t_r$ values. In the interpretation of the MC results, we
have  followed the  guide-line of accepting only the RN values
that do not present discrepancies between the largest $t_r$ data and
the asymptotic limit $E_{\infty}$. The fit (\ref{fit})
has been used to decide whether to trust or not  the MC values but
not to provide the asymptotic limit.

We have verified that at $\rho_0$,
and for values $ q<2.5$ \AA$^{-1}$, the RN energies using the Feynman
wave function (\ref{fey2}) do reach the expected constant regime, the
difference
between the largest $t_r$ calculated energy and the value of $E_\infty$
predicted by the
$\chi^2$-fit (\ref{fit}) being less  than the statistical error.
This is not the case for $\rho_P$. At this high density, that agreement
only subsists for the lowest $q$ value and for the value of $q$ nearest
to the roton. For the other values of $q$ we have had to include
backflow correlations (\ref{back2}) to reach the asymptotic
limit. This fact is illustrated in Fig. 1, where the
excitation energy
\begin{equation}
\varepsilon(q)= \frac{ \langle \psi_T(q) | H | \Phi(q) \rangle}
               { \langle \psi_T(q) |  \Phi(q) \rangle}
              - \frac{ \langle \psi_T^0 | H | \Phi^0 \rangle}
               { \langle \psi_T^0 |  \Phi^0 \rangle}
\label{epsilon}
\end{equation}
per particle is plotted as a function of $t_r$ for $q=1.11$ \AA$^{-1}$
and $q=1.84$ \AA$^{-1}$. Near the roton, $q=1.84$ \AA$^{-1}$, both the
Feynman and backflow results show a coincident asymptotic value without
a significant slope. At $q=1.11$ \AA$^{-1}$, near the maxon energy, the
situation is clearly different: the backflow results have reached a
constant behaviour whereas the Feynman ones show a slow approach to the
correct value, which has not been achieved yet for the maximum value of
$t_r$. The latter behaviour is also observed for the highest value of
$q$
($q \simeq 2.6$ \AA$^{-1}$) both at $\rho_0$ and $\rho_P$. In this case,
the inclusion of backflow correlations in the wave function is not
enough to eliminate the bias and a significant difference
exists between the largest $t_r$ energy and the asymptotic value
predicted by the numerical fit (\ref{fit}).

The released-node mechanism suppresses the fixed-node constraints and
drives the calculation to the exact excitation energy, as shown in Table
I. In the table, fixed-node values using $\psi_T^F$ and $\psi_F^{BF}$,
and the released-node estimation are compared with  experimental data
\cite{don} at the equilibrium density $\rho_0$. The FN results with
backflow correlations improve the
Feynman ones for the three values of $q$ in a magnitude which depends
on $q$. Thus, the inclusion of backflow correlations seems slightly more
relevant in the roton than in the maxon. On the other hand, the RN
excitation energies agree with the experimental data for the three
values of $q$ within the statistical errors.

In Fig. 2 the RN excitation energies are compared with the experimental
spectrum \cite{don} at $\rho_0$. The RN results correspond, for each
$q$, to
the last point in the release process, the error bars being only the
statistical errors. As commented before, the systematic errors are less
 than the statistical ones except for the highest $q$ result
($q=2.58$ \AA$^{-1}$). For this latter value of $q$ we also
report an estimation
coming from the extrapolation supplied by the fit (\ref{fit}). Apart
from this point, where the RN method shows the shortcomings of the
backflow wave function at a so high value of $q$, the agreement between
the RN results and the experiment is excellent. As a matter of
comparison, the FN results using $\psi_T^{BF}$ are also plotted. It is
worth noticing the difference between the FN energies in the maxon
and in the roton regimes; the roton is reproduced quite accurately
whereas in the maxon the backflow correlations overestimate
appreciably the excitation energies. At the highest $q$, where the
spectrum bends down, the FN energy is quite far from the experimental
data.

As is well known from neutron scattering data, the location
and depth
of the roton minimum depends on the density. Thus, when the density
increases the roton appears shifted to higher momenta and its energy
decreases. The energies in the maxon region increase with the density
but in an amount not so well experimentally known as in the roton. In
Fig. 3, we report the RN excitation energies at $\rho_P$ in comparison
with  experimental data \cite{woods}. There is, again, a good agreement
between
theory and experiment within the statistical errors except at the
highest
$q$ evaluated ($q=2.74$ \AA$^{-1}$) where the RN energy has not reached
a constant value inside the release interval.

In conclusion, we have shown that the diffusion Monte Carlo method in
conjunction with the fixed-node technique, and more specially, with
the released-node method provides a very useful tool to study
excitations
in correlated quantum many body systems like liquid $^4$He. The results
for $\varepsilon(q)$ are in an excellent quantitative agreement with
experimental data, both at the equilibrium and near the freezing
densities, improving previous variational and CBF results and providing
an exact description of one of the oldest and hardest problems
in the study of quantum fluids from a microscopical viewpoint. Possible
applications of the RN-DMC method would be the ripplon excitations in a
free liquid $^4$He surface or the determination of the excitation energy
of a single impurity in bulk liquid $^4$He. On the other hand, the
interpretation of the roton excitation as a single-particle mode
deserves further theoretical work  from a microscopical
viewpoint \cite{strin}.

We would like to acknowledge useful discussions with S. Stringari and H.
Glyde.
This work has been supported in part by DGICYT(Spain) Grant No.
PB92-0761 and No. TIC95-0429. We also acknowledge the supercomputer
facilities provided by the \'Ecole Polytechnique F\'ed\'erale de
Lausanne. J. C. also acknowledges a grant from Fundaci\'o Catalana per
la Recerca.

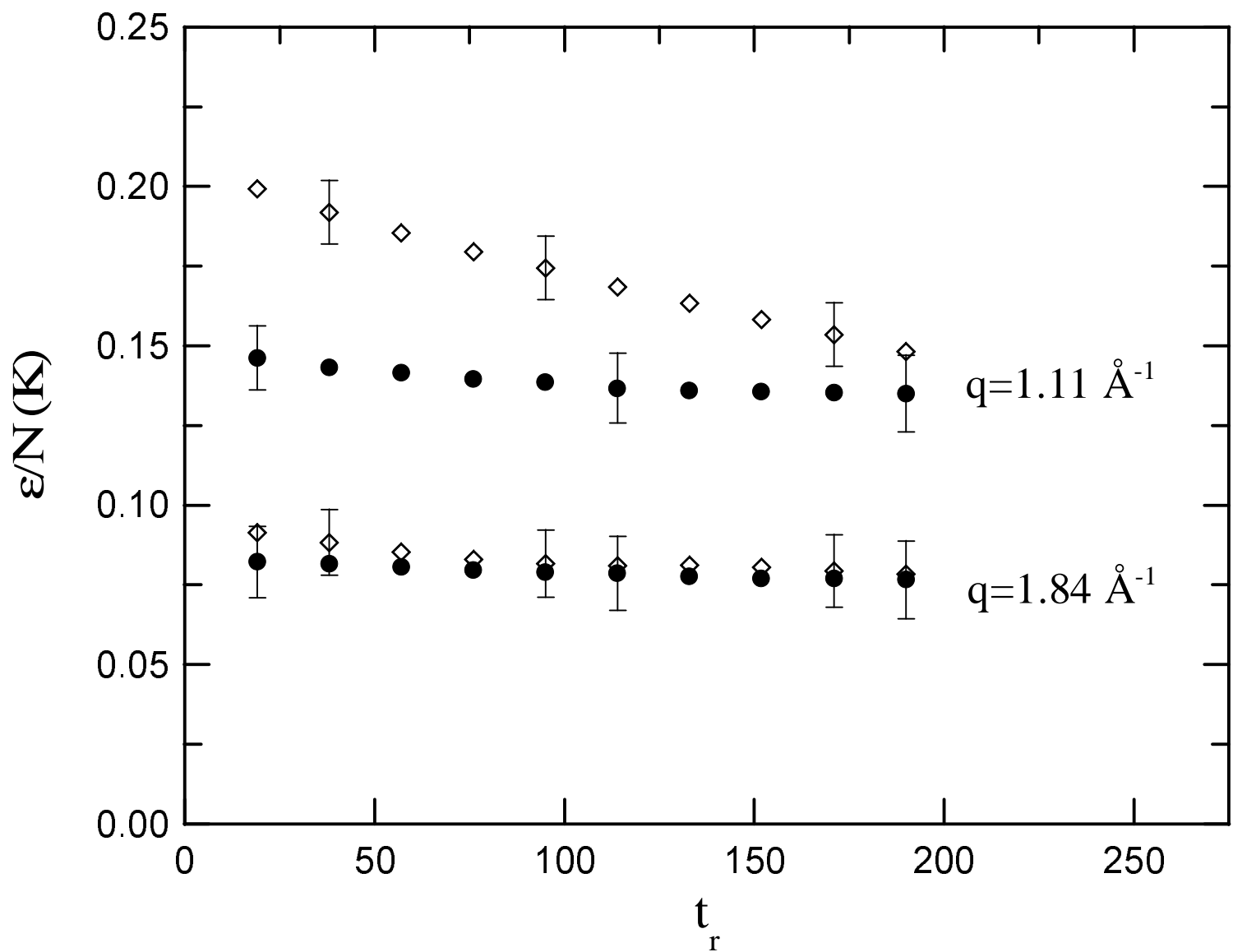
\begin{figure}
\caption{Excitation energies per particle as a function of the lifetime
$t_r$ at $\rho_P$. The full circles are
obtained using $\psi_T^{BF}$ and the diamonds using $\psi_T^F$.}
\end{figure}

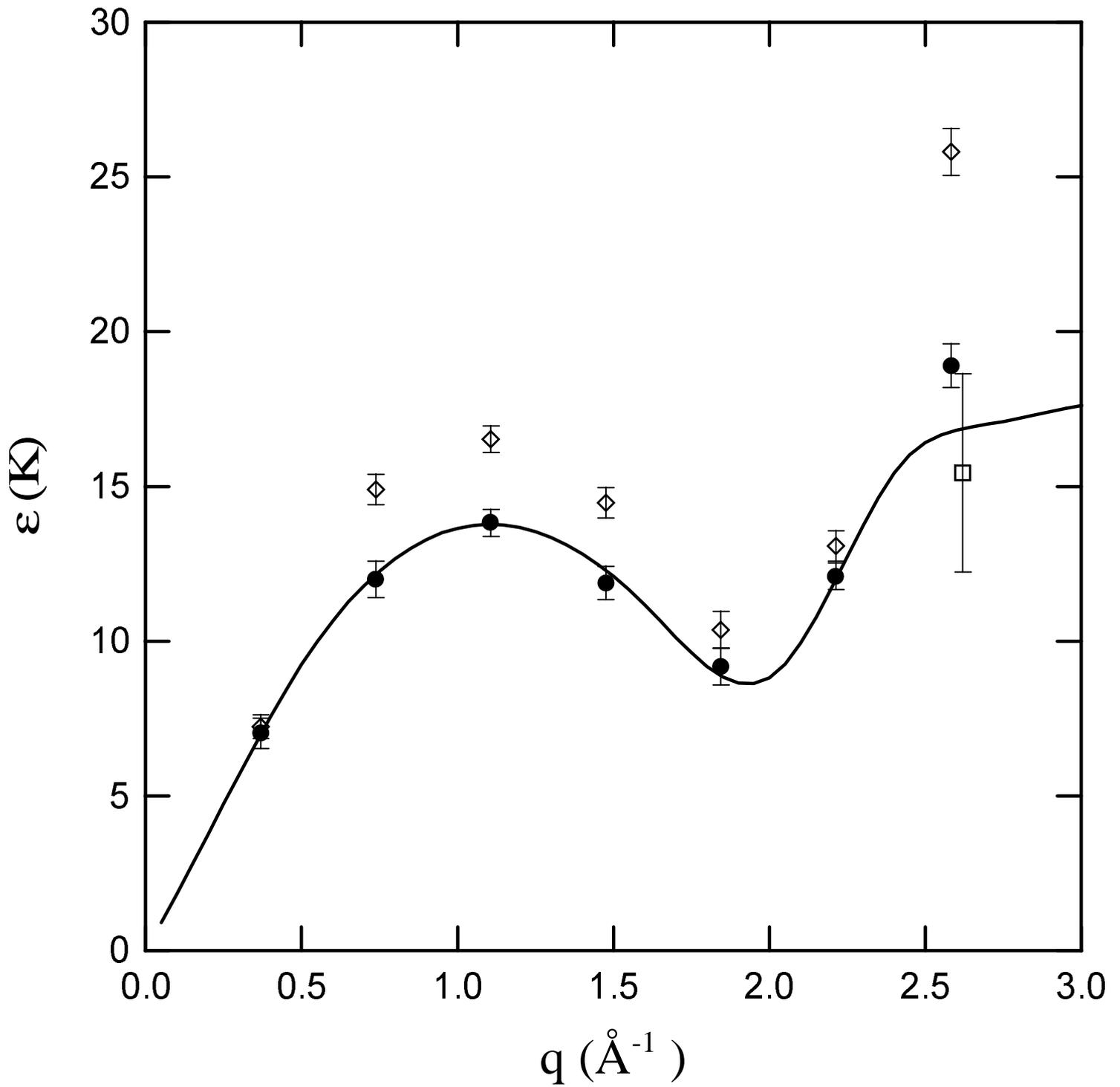
\begin{figure}
\caption{Phonon-roton spectrum at the equilibrium density $\rho_0$. The
full circles are the RN results and the diamonds correspond to a FN
calculation with $\psi_T^{BF}$. The open square, which has been slightly
shifted to the right for clarity, is the result of the extrapolation
with the fit
(\protect\ref{fit}). The solid line is the experimental data from Ref.
\protect\cite{don}.}
\end{figure}

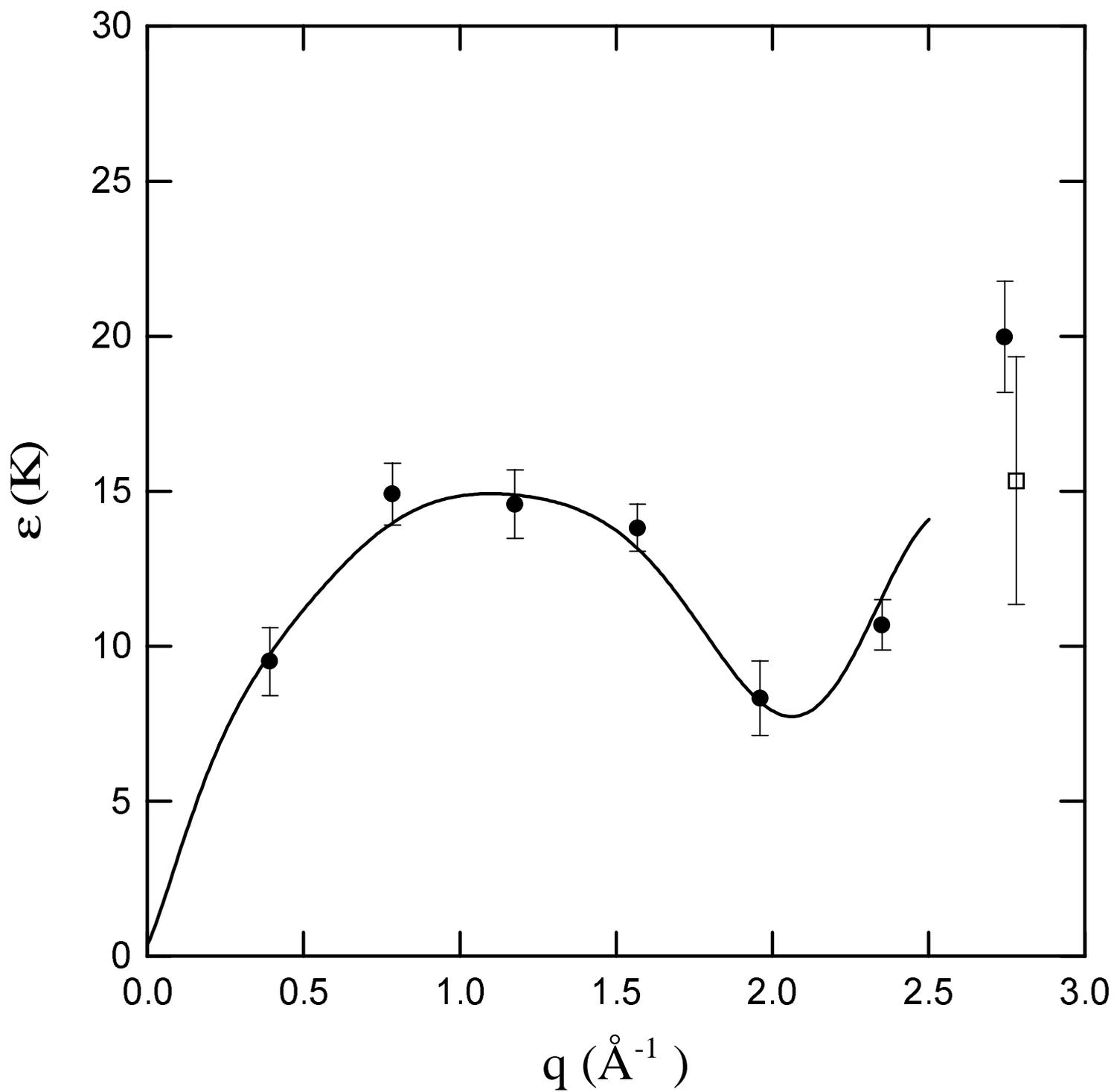
\begin{figure}
\caption{Phonon-roton spectrum at the density $\rho_P$. Same notation as
in Fig. 2. The experimental data is from Ref. \protect\cite{woods}.}
\end{figure}

\begin{table}

\caption{Excitation energies at $\rho_0$ in comparison with
experimental data. The FN-$\psi_T^F$ and FN-$\psi_T^{BF}$ columns are
the fixed-node energies using $\psi_T^F$ and $\psi_T^{BF}$,
respectively. The RN column corresponds to the released-node
estimation. Experimental data is taken from Ref. \protect\cite{don} }

\begin{tabular}{ccccc}
$q$ (\AA $^{-1}$) & FN-$\psi_T^F$ (K) & FN-$\psi_T^{BF}$ (K) &
RN (K) & Expt. (K)  \\ \tableline
0.369    &   7.56 $\pm$ 0.49  &  7.24 $\pm$ 0.38 & 7.02 $\pm$ 0.49 &
 7.0    \\
1.106    &   18.47$\pm$ 0.49  &  16.52 $\pm$ 0.43 & 13.82 $\pm$ 0.43 &
 13.8    \\
1.844    &   13.82$\pm$ 0.54  &  10.37 $\pm$ 0.59 & 9.18 $\pm$ 0.59 &
 8.9    \\
\end{tabular}

\end{table}

\end{document}